\documentclass[journal]{IEEEtran}

\usepackage{color}
\usepackage{epsf}
\usepackage{times}
\usepackage{epsfig}
\usepackage{epstopdf}
\usepackage{graphicx}
\usepackage{algorithm}
\usepackage{algorithmic}
\usepackage{amsmath}
\usepackage{amssymb}
\usepackage{amsxtra}
\usepackage{multirow}
\usepackage{mathtools}
\usepackage{subcaption}
\usepackage{multirow}
\usepackage{comment}
\usepackage[normalem]{ulem}
\usepackage{tabu}
\usepackage[dvipsnames]{xcolor}
\usepackage{multicol}
\usepackage{cite}
\usepackage{caption}
\begin{document}

\title{Towards Collaborative Mobile Crowdsourcing}

\author{
\IEEEauthorblockN{Aymen Hamrouni,  \textit{Student Member, IEEE},  Hakim Ghazzai, \textit{Senior Member, IEEE}, Turki Alelyani, \textit{Member, IEEE}, and Yehia Massoud, \textit{Fellow, IEEE}}\\
{\thanks {\hrule
\vspace{0.1cm} 
Aymen Hamrouni, Hakim Ghazzai, and Yehia Massoud are with the School of Systems and Enterprises, Stevens Institute of Technology, Hoboken, NJ, USA. (E\textendash mails: \{ahamroun,hghazzai,ymassoud\}@stevens.edu).\newline
Turki Alelyani is with the College of Computer Science and Information Systems, Najran University, Najran, Saudi Arabia.
(Email: tnalelyani@nu.edu.sa.)
}}\vspace{-0.0cm}}

\maketitle
\thispagestyle{empty}

\begin{abstract}
  
Mobile Crowdsourcing (MC) is an effective way of engaging large groups of smart devices to perform tasks remotely while exploiting their built-in features. It has drawn great attention in the areas of smart cities and urban computing communities to provide decentralized, fast, and flexible ubiquitous technological services. The vast majority of previous studies focused on non-cooperative MC schemes in Internet of Things (IoT) systems. Advanced collaboration strategies are expected to leverage the capability of MC services and enable the execution of more complicated crowdsourcing tasks. In this context, Collaborative Mobile Crowdsourcing (CMC) enables task requesters to hire groups of IoT devices' users that must communicate with each other and coordinate their operational activities in order to accomplish complex tasks. In this paper, we present and discuss the novel CMC paradigm in IoT. Then, we provide a detailed taxonomy to classify the different components forming CMC systems. Afterwards, we investigate the challenges in designing CMC tasks and discuss different team formation strategies involving the crowdsourcing platform and selected team leaders. 
We also analyze and compare the performances of certain proposed CMC recruitment algorithms. Finally, we shed the light on open research directions to leverage CMC service design.

\end{abstract}

\section{Introduction}
\label{Sec1a}
\textcolor{black}{The wide spread of mobile devices has enabled a new paradigm of innovation called \textcolor{black}{Mobile Crowdsourcing (MC)}. This emerging model has become a ubiquitous mean for individuals and organizations to use open-call format to attract and recruit many mobile users to complete specific tasks that require the use of the device's capabilities (e.g., built-in sensors such as GPS, accelerometer, and cameras as well as computational resources) along with the device holder's skills set (e.g., communication, judgment, expertise)~\cite{2}. By combining and leveraging the abilities of humans and the capabilities of machines, MC can accomplish tasks that are hardly handled by standalone IoT devices for a wide range of applications including weather reporting, health monitoring, traffic control, crowd and disaster management~\cite{8539015,8892766}.}

\textcolor{black}{In typical MC systems, there are three main components: task requester, worker, and platform. The task requester is an IoT user who crowdsources his/her tasks through an online platform. The worker is the potential contributor who will work on the posted task, using his/her own mobile/IoT devices, for different motivations, e.g., monetary or social rewards. The platform is the component responsible for managing the whole process and coordinating between the task requesters and workers.}

\textcolor{black}{MC is widely exploited in practice and spans many real-world use cases. As a first example, we cite the SenseCityVity project~\cite{3}, which is a MC service that uses the geolocalized images, audios, and videos collected by the citizens of Mexico to generate multimedia streams that enable a better understanding of the urban landscapes of cities in the Global South. Another MC example is the Fly-Navi project~\cite{9078872}, which consists in an indoor navigation system collecting WiFi sensory data from workers in order to correctly navigate the intended user towards his/her destination. Finally, we can cite the example of the event reporting MC framework that has been developed in~\cite{8982179} where nearby smart-phone users are recruited promptly and solicited to submit high-quality photos to cover ongoing events and keep track of any updates.}
 
In the aforementioned use cases, and in MC systems in general, the workers are participating together to address a given task but in an independent manner and without any sort of interactions. This form of collaboration in MC can be defined as \textit{silent collaboration}. However, with the proliferation of IoT technology and social networks, many more complex MC applications are emerging requiring a certain level of interaction and coordination among the workers to ensure the successful completion of the tasks. In this sense, selected workers equipped with assorted IoT devices can be regrouped together to form teams whose members are bounded to interact and collaborate together to complete different not necessarily homogeneous tasks. This team-based MC paradigm can be identified as \textcolor{black}{Collaborative MC (CMC)}.

%CMC involves the modern computing technologies to overcome some of the challenges that the traditional techniques can fail to tackle. 
By further boosting the interaction between workers, CMC complements IoT on a large spectrum of smart city applications that cannot rely on a basic data collection operation. For instance, community-based monitoring applications~\cite{8355907}, which are defined as the process where concerned community groups collaborate together to monitor, track, and respond to issues of common environmental concerns, is one of the prominent applications where CMC can be very essential. Other complicated CMC projects, such as evacuating hazardous areas with a high spatial coverage and without the need for deploying a dedicated sensor network, can require the intervention of multi-disciplinary workers equipped with diverse IoT devices (e.g., smartphones, CCTV, UAVs). In transportation, CMC can be useful in mitigating GPS outage in network blocking environment such as tunnels. A team of workers (e.g., vehicles, smartphones) could be collaborate to estimate the GPS positions inside the blocked areas. Devices at the entrances can provide their GPS signals and share it with other devices that will estimate the distances separating them, e.g., dead reckoning. Other devices with high computational resources can help in running the localization algorithms.

\textcolor{black}{Depending on the CMC project and after providing explicit and informed consent, workers can be asked to complete what is necessary in a participatory and/or an opportunistic way. Participatory CMC systems require informed consent from the workers each time in order to conduct a manual intervention to complete the task. In contrast, opportunistic CMC systems rely on implied consent to collect and report the data from the workers' IoT devices automatically.} Many issues may arise when addressing the very complex nature of collaboration in CMC from technical and social perspectives. It is mandatory to ensure well-connected, socialized, and effectively coordinated workers to successfully execute the posted tasks. Also, the design of CMC systems needs to account for the technological complexity such as scalability, processing, and quality of service.

This study delves into the CMC paradigm and its applications. It provides a detailed taxonomy of the model and investigates a number of challenges that needs to be addressed in the context of CMC. Our study encompasses mobile "crowdsourcing" tasks in general rather than mobile "crowd-sensing" to emphasize that CMC projects are beyond the basic sensing and data collection tasks. We also present different strategies for forming effective and socially connected teams composed of IoT workers considered as an essential criterion for successful CMC projects. We support our study with Monte Carlo simulations where we: i) compare three team formation strategies depending on the entity responsible in recruiting the team members: the CMC platform, a team leader from the crowd, or a hybrid approach involving both, and ii) investigate the performances of the first recruitment strategy using different recruitment algorithms. At the end, we discuss open research directions that are vital to maturate CMC systems in practice.

\section{CMC Work-flow and Taxonomy}
\label{workflow}

\begin{figure}[]
\centering
     \includegraphics[width=9cm]{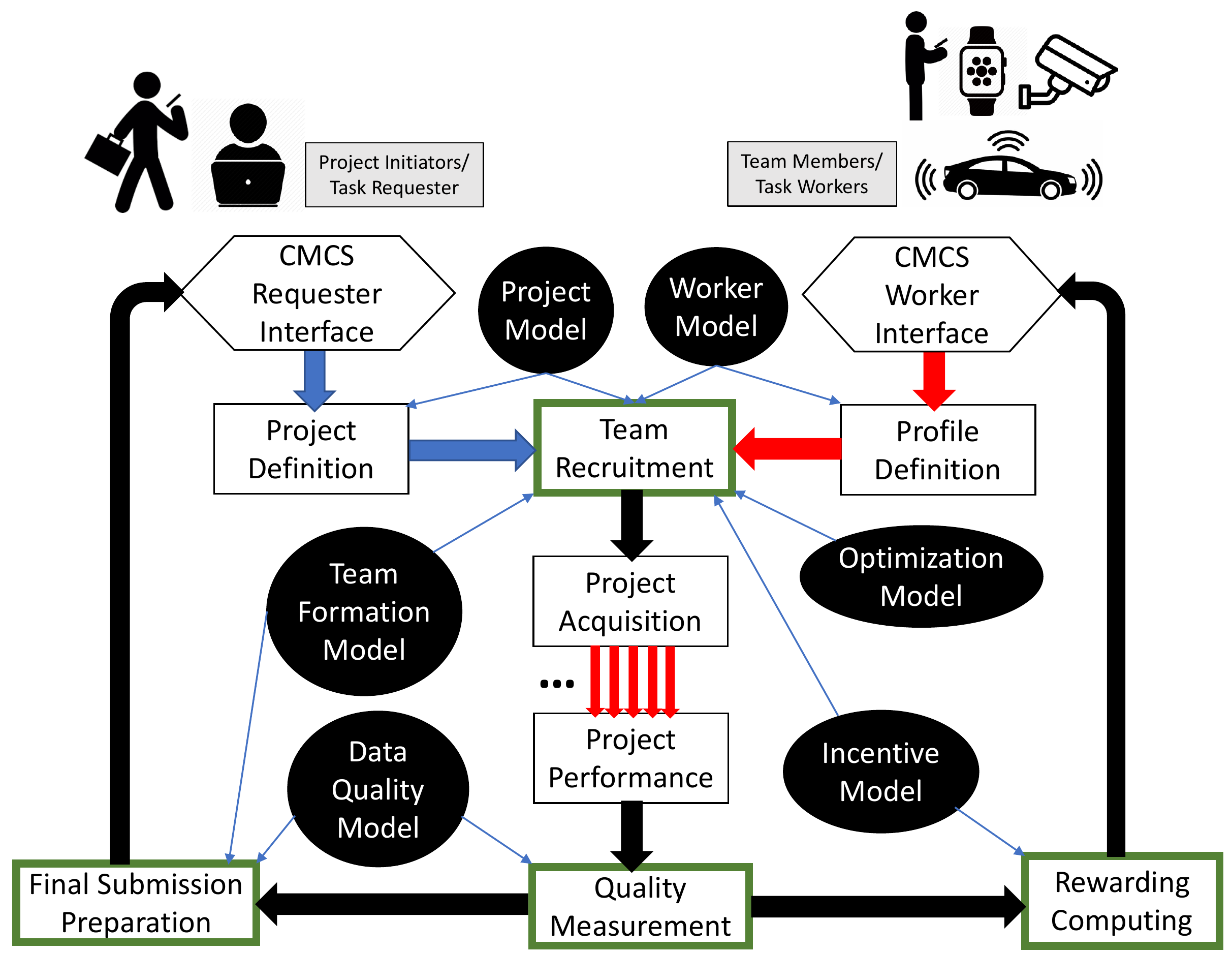}
\caption{\textcolor{black}{General work-flow of a CMC system. Black circles represent the CMC system models. The green boxes represent the actions of the platform. The blue, red, and black arrows represent the actions conducted by the task requester, workers, and platform, respectively.}}
\label{cmcsworkflow}
\end{figure}

A CMC system consists of a central platform residing in the cloud, a huge number of users owning IoT devices (e.g., smartphones, laptops, or other smart machines) acting as workers and connected to the platform~\cite{9125433,9125426}, and project requesters, that can correspond to other IoT devices' owners as part of the system or another external entity, e.g., a local authority. When in need, project requesters submit a complex task or a project to the platform in order to hire skillful and efficient teams of workers to collaboratively execute them. \textcolor{black}{A generic simplified work-flow of the CMC system is illustrated in Fig.~\ref{cmcsworkflow}. Once workers' and tasks' attributes are defined, the CMC platform proceeds with matching suitable teams to projects and tracking the status and progress of the active tasks. Once a team is selected and upon receiving notifications from the platform, crowd participants start to collaboratively work together to execute the task, e.g., by sensing data or providing computational resources through their smartphones.} This collaborative model with geographically distributed smart devices can provide on demand large-scale services by stimulating crowd participants to exploit, join, and share their devices' resources and features. 

\begin{figure*}

\hspace{-0.5cm}
\centering
     \includegraphics[width=19cm]{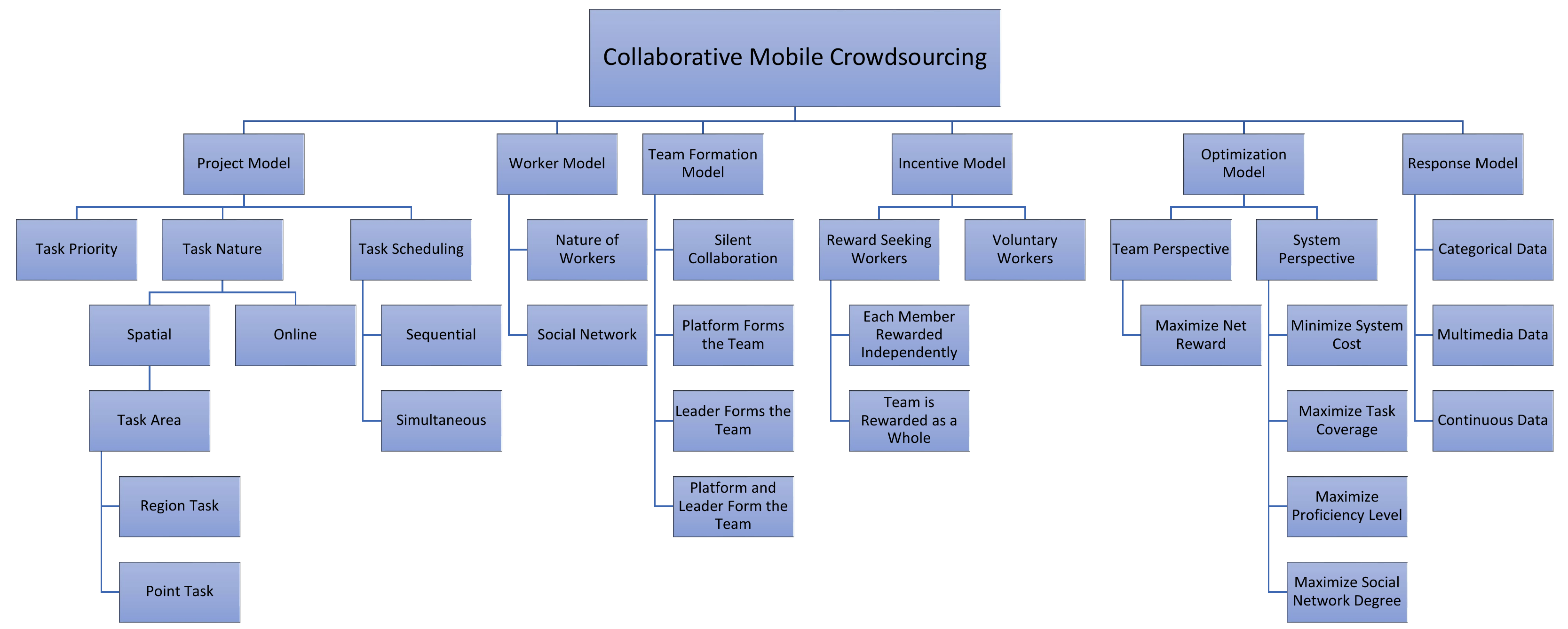}
\caption{\textcolor{black}{Taxonomy of CMC systems composed of six main components: the tasks/projects model, the worker model, the team formation model, the incentive model, the optimization strategy model, and the response model.}}
\label{taxonomy}
\end{figure*}

The effectiveness of crowd collaboration while simultaneously submitting the results have been supported by previous studies as a way to improve performance and increase scalability. However, this novel format increases the complexity to manage the entire process including but not limited to participants motivation, design mechanisms, coordination, incentives, and quality assurance. Therefore, the success of these types of systems requires a systematic design that would take into account the level of complexity of the proposed project. To assist on identifying the needs of future applications, we have developed a taxonomy for CMC systems. As depicted in Fig.~\ref{taxonomy}, the first division relates to how CMC tasks/projects are modeled. The second division is dedicated to the worker model to identify its skills, the type of owned IoT devices and its social relations in the network.  The third division deals with the types of team formation strategies that can be adopted in CMC and that we will discuss in detail in Section~\ref{teamformation}. Incentive classes requested by the workers are presented in the fourth division while the fifth division illustrates what a CMC application aims to optimize and what constraints it may face. The sixth division relates to various types of responses provided by workers, they can be either categorical, e.g., answering a yes or no question, continuous, e.g., a physical input measure, or multimedia, e.g., photos or video streams.

\section{Design Challenges of CMC}
\label{CMCSDESIGN}
In this section, we elaborate on the main challenges of designing robust CMC systems.  

\subsection{Project and Task Design}

\textcolor{black}{Different CMC systems offer various types of tasks where crowd workers may have the choice to complete what is required in a participatory or opportunistic way. There are several characteristics that MC and CMC have in common and which represent the main components in task design. For example, factors such as appropriate user interface, task presentation, and wording style suggest that different design patterns should be adapted depending on the task type and worker group. Also, coherent protocols design and suitable computing interfaces defining the interactions between multiple software intermediaries must be carefully studied in order to ensure communication smoothness between IoT workers. On top of this, the CMC platform needs to evaluate the available resources including storage, power, and computational capabilities that are required to perform the task. }

On the other hand, designing projects in CMC must consider several other challenges. Task management design is the most important one. When managing CMC tasks, the following features form the most important aspects: i) task complexity, ii) task  \textcolor{black}{workload distribution}, and iii) time.
The task complexity, related to the cognitive dimensions of the CMC task design, must be clearly distinguished from the task difficulty, which concerns the worker's perceptions of the task itself. Nonetheless, certain CMC tasks can be difficult and require high effort without being really complex just because of their confusing task design~\cite{10.14778/3067421.3067431}. 
The CMC platforms need to design a general and convenient task sub-structure that ensures a clear distribution of the workload among the team members and avoid any conflicts or misguidance that may occur. Moreover, defining explicit boundaries between the roles of team members is a key feature in CMC and the quality of the resultant project heavily depends on it. Time is another CMC challenge in task management. For instance, the appropriate time to post a task and the time needed to submit a task play major roles in crowd performance. 
The task design must cover the availability time of each worker in order for the CMC framework successfully manages the assignment process and recruits team members that are simultaneously available to complete the project with well-defined schedules when needed.

%To sum up, the CMCS platform needs to provide ultimate conditions for the requester, guide him/her to precisely define the task, and avoid any complications. The task requester, on the other hand, needs to be clear when defining his/her desired task and uses the platform’s provided tools and attributes to sharply specify his/her needs. 

\subsection{Incentive Mechanisms}
\textcolor{black}{The workers' rewards process, \textit{aka} crowd rewarding, can be either achieved through extrinsic or intrinsic incentives\footnote{\textcolor{black}{The platform is aware about the participation of workers in different projects and rewards them according to their contribution.}}. Crowd workers are mostly driven by the extrinsic reward they receive for their work effort and any other resource consumption (e.g., device's energy utilization or worker traveling cost).} Therefore, underestimating or overestimating the price may affect the quality of the contributions. 

There are several studies that proposed different strategies for incentive design in classical MC, e.g., using reverse auction, distributed truthful rounds, and tournaments~\cite{9043743,7875128}. In all of these approaches, there is no guaranteed "up-front" amount, where workers are partially rewarded before task completion, and the whole reward is claimed only after completing what is necessary. Some scenarios suggested that workers are paid according to the quality of their contributions. Other ones suggested that workers are advised before their operation that their rewards are only approved upon the successful job completion. Workers who provide partial or wrong submission would not be compensated. %Incentive mechanisms based on reverse auction, distributed truthful rounds, and tournaments, have been proposed in the literature while considering several factors such as quality of submissions, fairness among workers, and task requester budget.}

\textcolor{black}{
However, most of these MC incentive mechanisms are not applicable in the collaborative context. One of the major challenges that can be encountered in CMC rewarding is the effort needed to be done in case there is a large pool of workers. Moreover, in the collaborative context, this may not be the optimal design as there would be a number of teams and each team has a number of workers which would require more coordination effort. Possible CMC rewarding approaches can include either: i) rewarding the whole team upon successful completion as a whole or ii) rewarding each team member separately based on his/her contribution independently of the other team members. The problem with the second approach is that the project requester will provide rewards even if the project was not completed. Other approaches suggest either providing weighted incentives where workers are rewarded based on their actual contribution or equally distributing rewards among the team members.}

\textcolor{black}{
The other possible CMC form of incentives is the intrinsic rewards, or sometimes referred to as social rewards. These rewards represent the satisfaction that the worker gets from the project itself, such as having pride in his/her own work, showing personal growth, feeling respected by other team mates, and being part of a team.}%Sometimes, a worker may be willing to reduce a part of his/her extrinsic reward (e.g., their monetary reward) in exchange of being able to work among his/her friends or preferred colleagues. This approach can be beneficial for both, the worker and the project requester, because the worker, surrounded by his/her friends, will exert less effort to communicate and hence, will be willing to forgo some percentage of his/her initial earnings. }

\subsection{Team Formation}
Team formation is at the core of this type of mobile crowdsourcing. \textcolor{black}{The teams are formed when needed in accordance to the projects' requirements and their availability.} Recruiting a group of workers to collaboratively focus on crowd tasks is complex and requires rigorous effort especially on how the team can be formed. First, in order to the platform to recruit workers and successfully coordinate between them, it is important to determine the team size, the team leader, and the responsibility of each member. Second, after recruiting each worker, the platform is given the option to form a number of teams and assign leaders to them. There would be a hard deadline where all teams should be formed and their results are returned to the platform. Third, team members may not be required to work with each other simultaneously in order for the task to be submitted. For example, the output of worker "A" requires the output of worker "B" as an input. 

It is worth to note that task requesters may choose to recruit more than one team for their CMC projects. Hence, the platform needs to efficiently manage the available workers satisfying the skills requirement for each of these projects. Indeed, the platform is constrained by a maximum number of teams that depends on the possible combinations of workers available at a certain time instant. Many other factors should be considered to ensure ideal teams in CMC including age, gender, location, and owned IoT devices' specifications. Different recruitment strategies will be investigated and evaluated in Section~\ref{teamformation}.

\subsection{Response Submission and Quality Monitoring}
In the typical MC, the data returned is checked and validated for errors and misleads before being forwarded to the task requester. However, proceeding with the same technique in CMC systems is not straightforward. In fact, the CMC collected data needs to go through a process of synchronization, during which the submitted solutions from different team members are collected and processed. Moreover, the need to combine the workers' efforts in CMC increases the error margin which in turn decreases the final quality submission.

Quality is a long-standing issue in MC systems in general. As crowd workers come from very diverse backgrounds and possess devices with different specifications, continuous process in assessing the system quality is very important. One way to estimate the quality of the future worker submission is to assess beforehand his/her attributes (e.g., degree of expertise) and his/her device resources before assigning him/her any task~\cite{Chen2018}. However, this proactive approach showed over the time low efficiency. Another possible approach is reactive, which evaluates the worker's submission after hand and the quality monitoring process iterates through the workers' submission phase (i.e., the worker keeps submitting task result until the platform decides that the quality is acceptable). Another aspect on ensuring a high quality in CMC is the communication channels between the recruited team members. In fact, guaranteeing the proper interaction between workers and platform management can lead to the satisfaction of the team members which in turn leads to better quality. 

%Appropriate strategies should be applied according to which context a task falls into. Recent studies in MC have been introducing clustering algorithms powered by machine learning capabilities that allow reasonable data collection to be occurred~\cite{2}. Nevertheless, CMC remains in its infancy when it comes to data integration that assures scalability and quality. 

\begin{figure}
\frame{      \includegraphics[width=8.75cm]{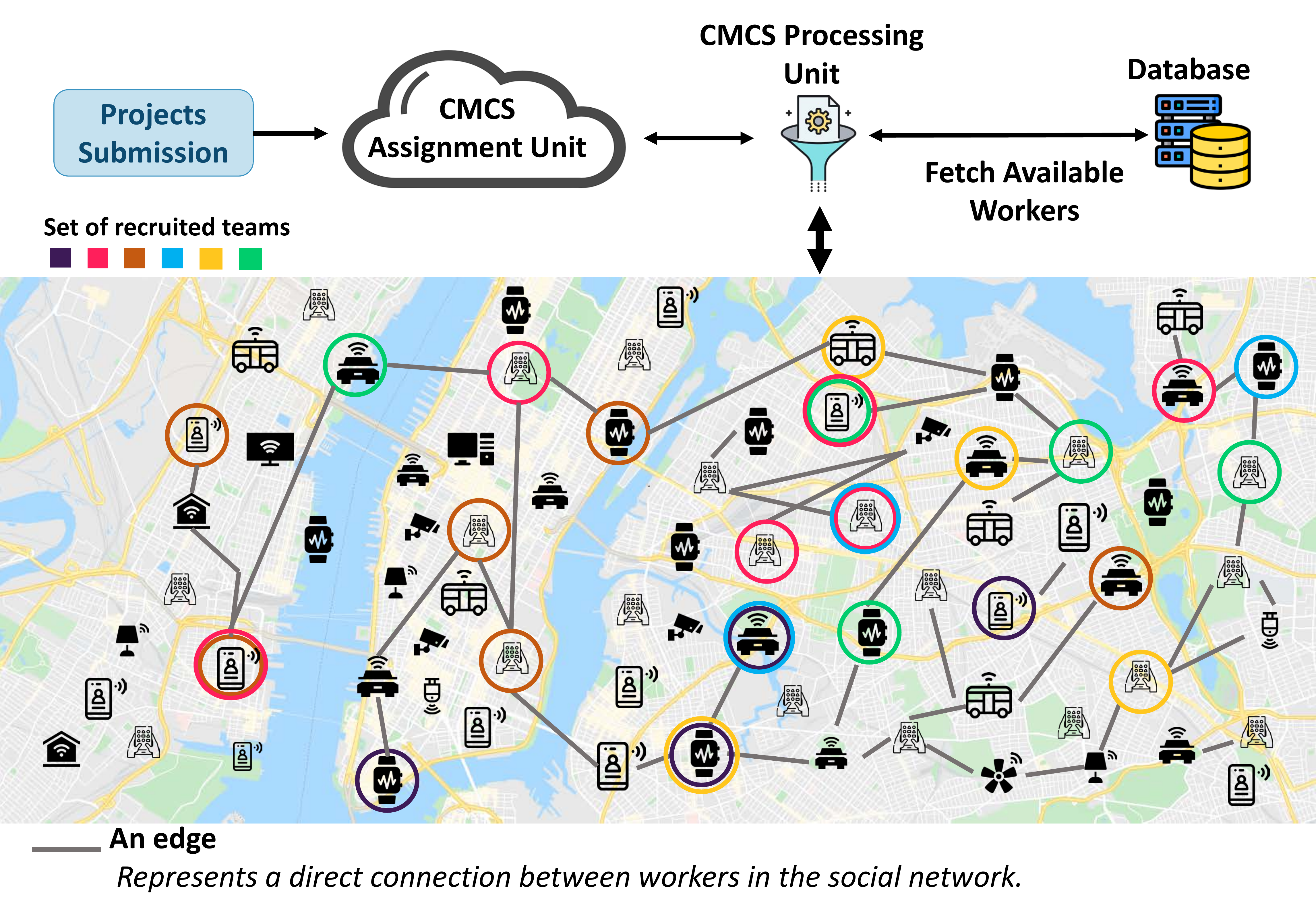}}
\caption{A map depicting the team formation process in the CMC context showing the workers' positions and their corresponding teams. Team formation is based on several types of IoT users recruited together to cover a set of tasks/projects. }
\label{visualization}
\end{figure}

\subsection{Optimization Goals and Constraints}
%In a CMC system, there are different possible approaches on how to optimize the recruitment and team formation processes.

From a worker's perspective, the goal is often to maximize the total net reward: the difference between the reward he/she gets from the system and the cost (e.g., depreciation cost). To achieve this goal, a worker may seek to be part of as many teams as possible to be selected even if they are competing with each other. This can be formulated using different game theory models to achieve Pareto optimality.

From the platform's perspective, the goals are often to maximize the CMC task successful completion rate with least cost and obtain maximum quality. In short, these goals can be summarized as follows: \\
$\bullet$ Maximize  task  coverage: This is to maximize the number of assigned projects. To achieve this goal, the CMC server first collects all the data of the available workers and then devises a strategy to maximize the overall number  of  assigned  projects.\\
$\bullet$ Minimize system cost: The  cost can be defined as the total incentives paid to each selected team member. Some systems may set their goal to hire a team with the lowest cost possible.\\
$\bullet$ Maximize  data  quality: Depending on how data quality is defined, different strategies can be applied to maximize data quality.\\
$\bullet$ Minimize number of tasks with missed deadlines: CMC tasks may have time constraints, so the selected teams need to complete the project   before the deadlines. In this case,  the system may want to minimize the number of projects that are not completed within the deadlines.  
\begin{figure*}[t]
\hspace{-0.6cm}
\subfloat[Team Skill Level]{     \includegraphics[width=4.4cm]{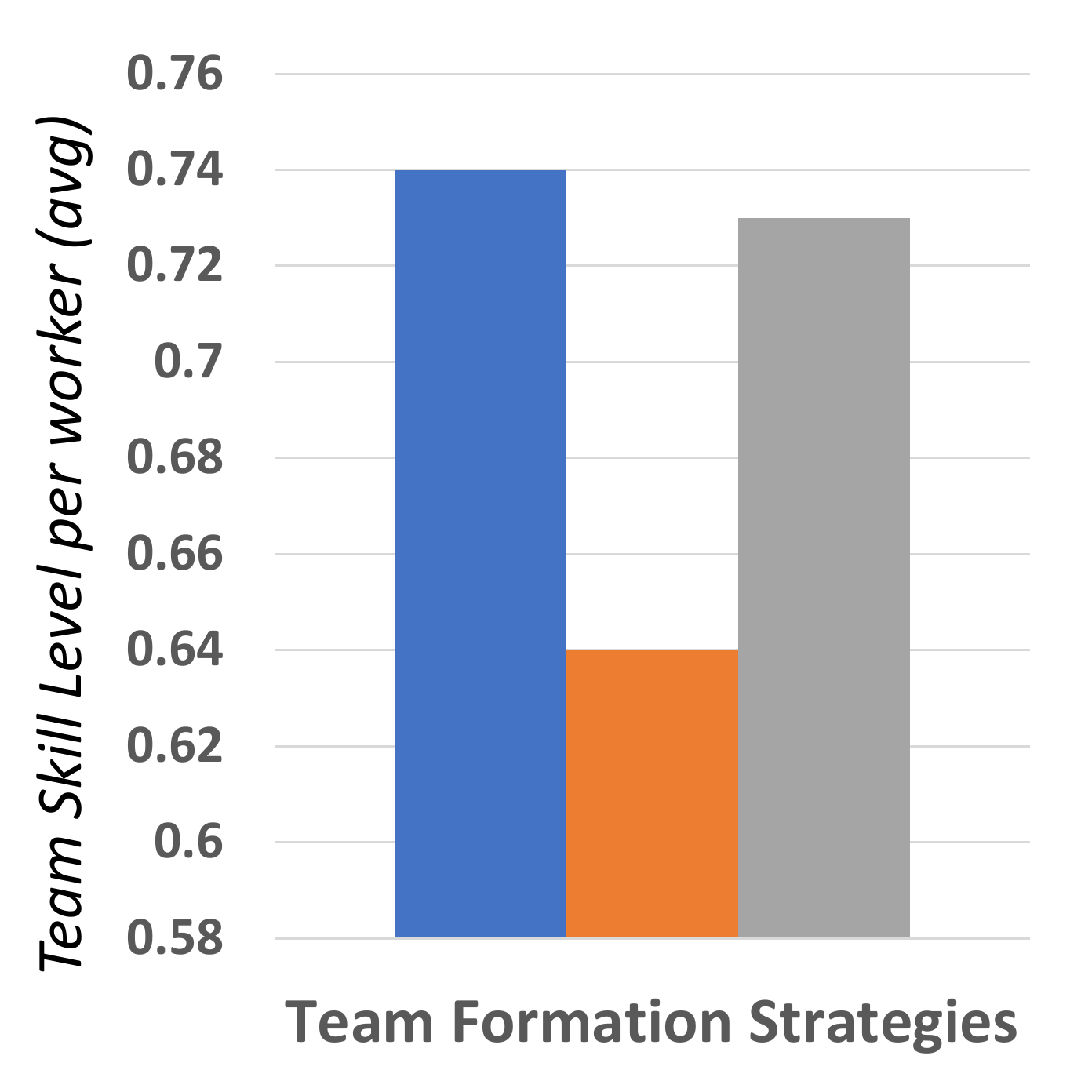}
}
\subfloat[Relationship Level]{ \hspace{-0.55cm} \vspace{0.1cm}\includegraphics[width=4.3cm]{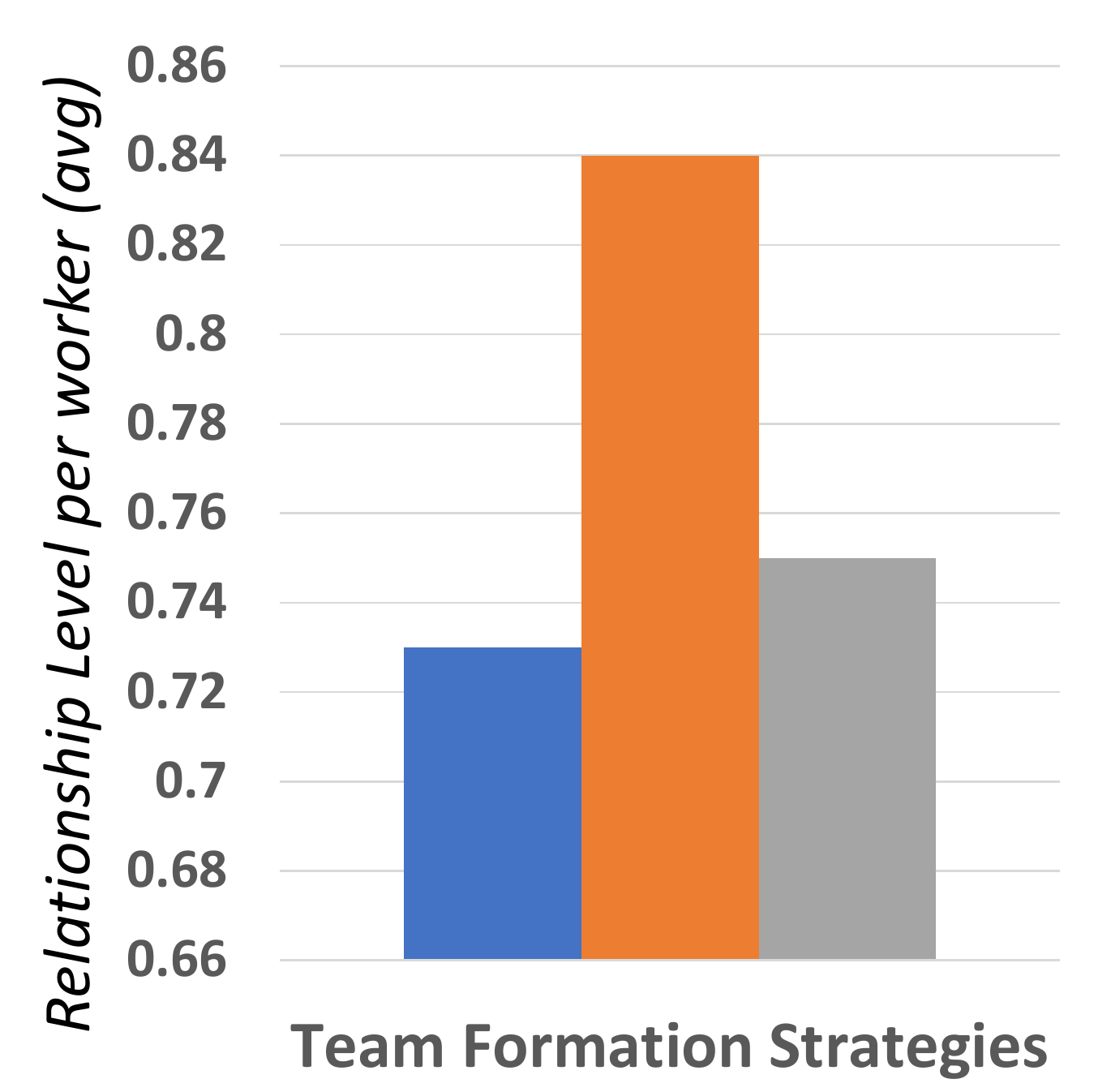}}
\subfloat[Recruiter Uncertainty Level]{  \hspace{-0.55cm}    \includegraphics[width=4.6cm]{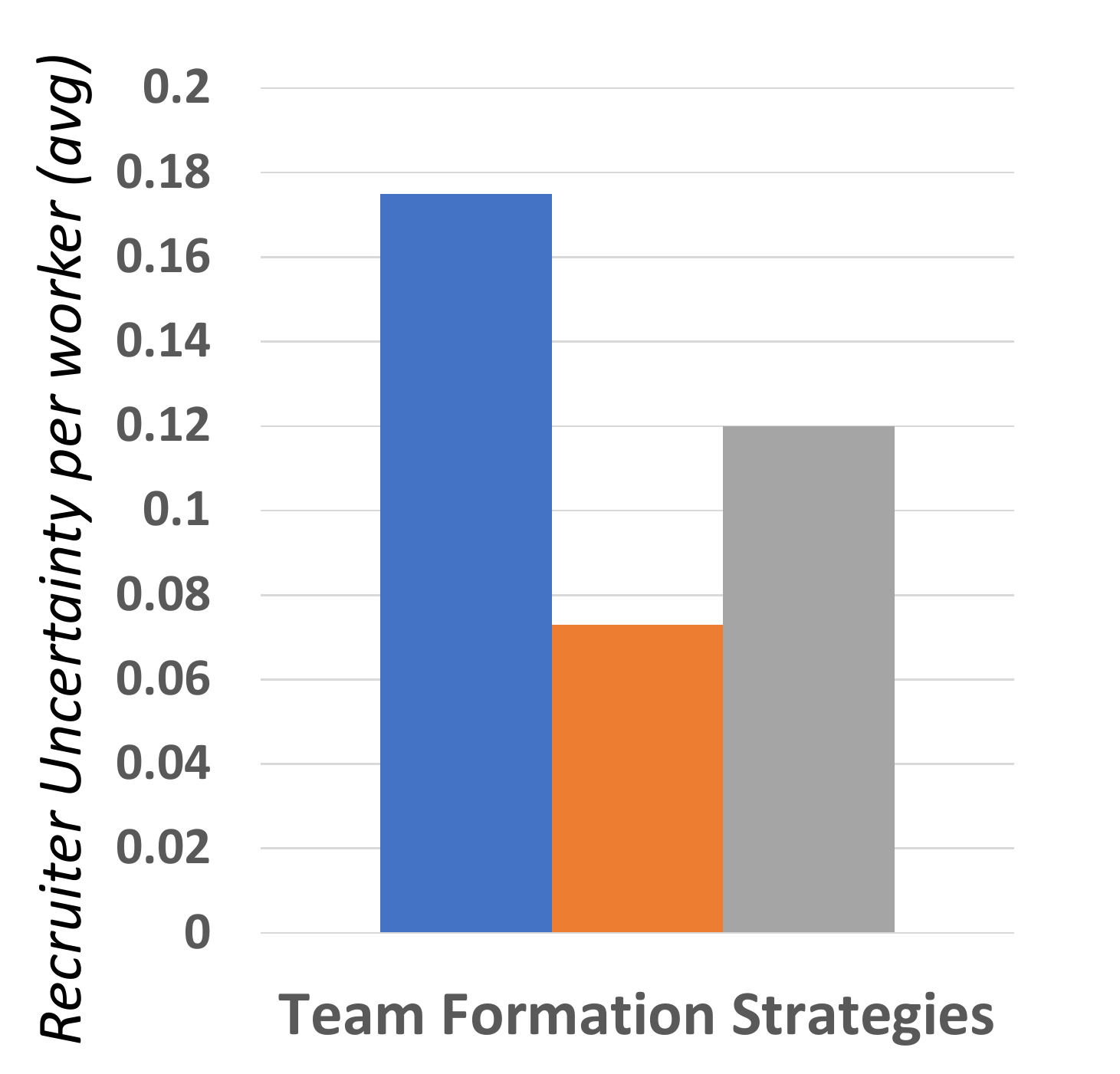}
}\hspace{-2.6cm} 
\subfloat[Team Cost]{\hspace{1.9cm} \vspace{-0.05cm}\includegraphics[width=6.4cm]{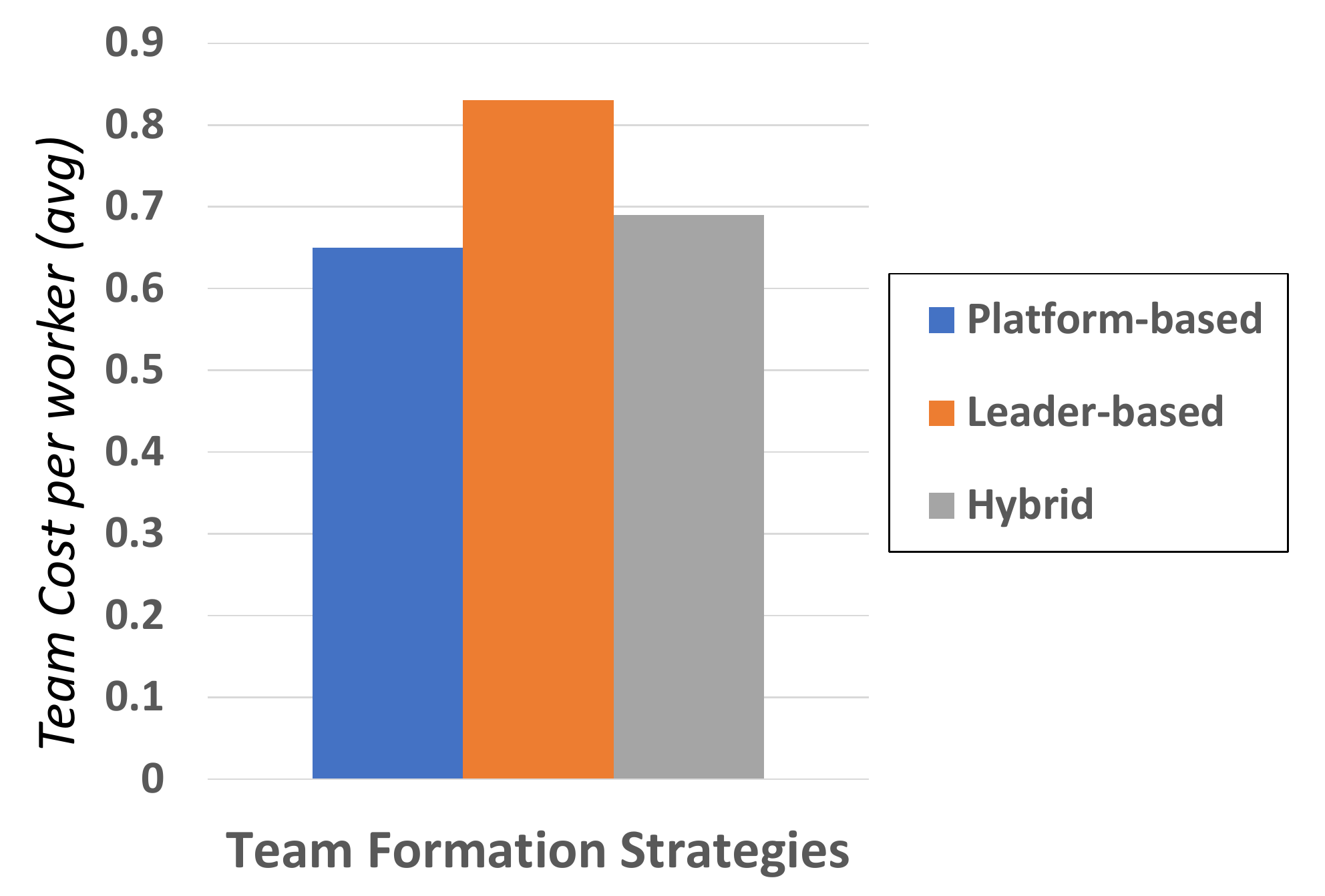}}
    \caption{Recruiter uncertainty level and recruited team skills vs. social network level for both, the leader-based and the platform-based strategies with the number of workers is equal to $20$ and the number of required skills is set to $7$.}
  \label{4figures} 
\end{figure*}
\begin{figure}[t]
\hspace{-0.3cm}
      \includegraphics[width=9cm]{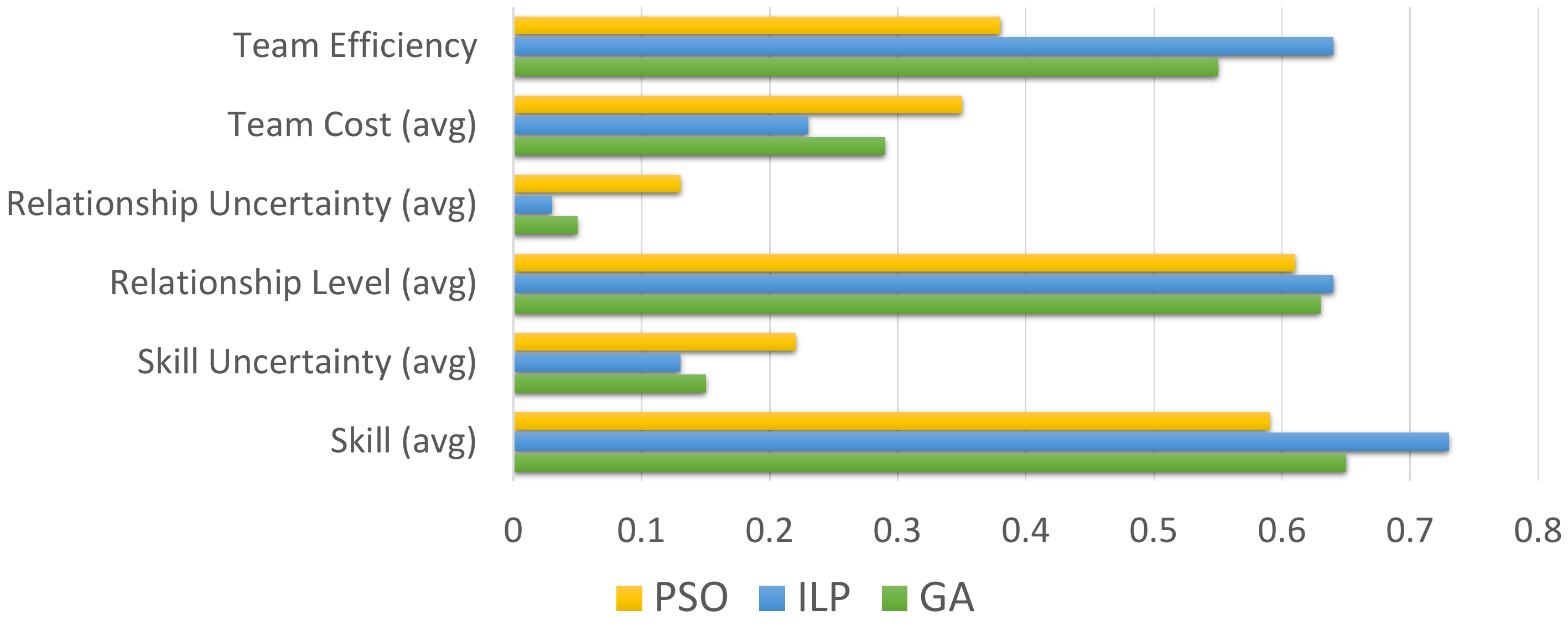} 
\caption{Performances of a CMC Platform-based strategy using three different algorithms (an optimal ILP solution) and two meta-heuristics showing the resultant values for the skills levels, skills uncertainty, social relationship strength, social uncertainty, and cost where the number of workers is equal to $20$.}
\vspace{-0.3cm}
\label{perfor}
\end{figure}

\section{Team Formation Strategies in CMC}
\label{teamformation}
%In this section, we focus on the most challenging part in CMC systems, namely forming suitable teams using the available workers and matching them to most convenient projects. 
In order to ensure an effective recruitment in CMC, it is very important to consider, in addition to the workers' skills, their social connections. As shown in Fig.~\ref{visualization}, workers are spatially distributed and connected through a social network graph, in which the nodes correspond to all the workers in the platform while the weighted edges represent the relationship level between them. The nodes in the social network represent the device's owner even in the case of an autonomous IoT device (e.g., CCTVs, UAVs). The CMC recruiter needs to discern the workers' features as well as the structure of the social graph to hire teams with high chemistry level and hence, increase the overall chances to successfully complete projects. %Unlike the classical MC systems
%, workers are recruited independently of each other to collaborate without the need of knowing each other or interacting together to complete tasks
%(i.e., silent collaboration). 
%We exclude this form of collaboration from our study since it does not reflect the nature of collaboration in CMC systems. 
In fact, collaboration strategies in CMC surpass the typical silent collaboration form of MC by incorporating the social relations as a criterion of selection. Thus, the formed team is, not only skilled, but also socially connected.

\subsection{Platform-based Strategy}
In this form of collaboration, the CMC platform is responsible of recruiting the entire team and matching it to the convenient project based on its own knowledge given the workers' attributes (e.g., profile, history, experience, previous performances, reliability). Moreover, the platform may establish a database of its registered workers where it could keep a log of all of their previous activities, interactions, social relations, and areas of expertise. Based on all of this knowledge, the platform forms the entire team and affects it to the project. \textcolor{black}{ This strategy is favorable to be adopted when the platform has sufficient history about the available workers and their previous collaboration. In other words, opportunistic systems are great candidates for this strategy since workers have more chance to build a long history log in the CMC platform. As a use case, we cite the example of mitigating GPS outage in network blocking environment which can be performed multiple times automatically without the need of manual intervention.}

\subsection{Leader-based Strategy}
This strategy suggests that the recruiter of the group is a chosen leader delegated by the platform to recruit the rest of the team. The recruiter, unlike the previous strategy, is a worker and the recruitment process is based on his/her knowledge. To select the leader, the platform may consider different criteria such as selecting the oldest, the most experience, or the highest socially connected worker. In the leader-based strategy, the recruiter does not only hire the team members but also monitor and coordinate the project among all the workers. For this strategy, there are two variations: i) the leader is also a team member with a required skill, and his/her role is also to contribute to the task by fulfilling a required skill, or ii) the leader role is restricted to hiring the suitable workers, supervising, and monitoring the project completion. \textcolor{black}{Each of these variations can be suitable for different projects. For example, if a project requires strong coordination and precise supervising, the leader needs to be more focused on that. Also, one variation can be favored over the other according to different factors, including, but not limited to, the nature of the project itself, the quality of the team members, the characteristics and the performances of their devices, etc. Therefore, the recruiter may decide the best suitable option based on its overall knowledge.  The leader-based strategy is more likely to be adopted when the social network connecting the available workers is dense. This can be the case of the CMC platforms for search and rescue activities where workers are co-located and are more susceptible to know each other, especially the team leader.}

\subsection{Hybrid Strategy}
This strategy combines both, the platform and workers knowledge to recruit the team. The process is as follows: the platform delegates a worker to recruit a suitable team based on his/her knowledge about the social network in his/her vicinity. Then, it uses its own knowledge to validate the leader choices. If the platform finds better options, it notifies the leader with its recommendation. Otherwise, it approves the leader choices. This strategy combines the local knowledge of workers and the global knowledge of the platform. %Because the noise levels on the attributes of workers in the leader-based strategy are propositional and increase with the number of intermediary workers between the recruiter itself and other workers in the social network graph, the platform may have better knowledge than the leader in socially distant workers.
%In this strategy, the platform can either approve the choices of the team leader or provide recommendation to the team leader. 
Other possible approaches may combine the platform and leader knowledge when choosing the workers. However, although they could yield the most efficient outcome, they require longer negotiation phase before both agreeing on a final decision which may cause sever delay in the team formation process. \textcolor{black}{ This strategy is more applicable for CMC projects involving both opportunistic and participatory tasks.}

\subsection{Performance Evaluation of Different Team Formation Strategies}
\textcolor{black}{In this section, we conduct high-level experiments on the three discussed earlier recruitment strategies: the platform-based, the leader-based, and the hybrid (Fig.~\ref{4figures}). Also, we perform performance analysis on three recruitment algorithms: Genetic Algorithm (GA), Particle Swarm Optimization (PSO), and Integer Linear program (ILP) applied to the platform-based strategy (Fig. \ref{perfor}).}

In order to simulate the team formation process for the three mentioned strategies, we formulate them as ILPs~\cite{hamrouni2020optimal} where our goal is to maximize a fitness function containing four key recruitment metrics: \\
\textcolor{black}{$\bullet$ Team skills efficiency: This metric measures the degree of expertise of the selected team calculated using the average skill level of all the team members. The higher the value of this metric is, the more skillful the team is. \\
$\bullet$ Team cost: it considers both, the total monetary reward requested by the team members, and other potential service fees (e.g., cost of traveling to the task location), resulting in the total cost that the task requester will have to pay if the team has been recruited.\\
$\bullet$ Team social relationship strength: This metric measures the relationship degree in the social network between the recruited team members. 
The team relationship level highlights the interactions between the workers within the selected team and describes indirect relations between the workers in the social network. High values of relationship levels within a team signify that the selected workers are more familiar with each other.\\
$\bullet$ Recruiter uncertainty level: it reflects the uncertainty of the recruiter towards the workers' attributes and describes the error distribution. As mentioned earlier, since the recruiter has only partial knowledge about the worker's attributes as it is highly susceptible to error, it is only fair to introduce noise to the workers' attributes. This error can be modeled as a normal distribution. The higher the confidence level (i.e., the lower the uncertainly error) is, the more likely that the chosen team members' attributes are close to the recruiter's expectations. } 

\textcolor{black}{
We consider a synthetic data with different types of project requirements and workers' skills. We use the Watts\text{-}Strogatz network model to create workers' social network graph and randomly produce their distribution with small-world properties. The skills and the relationships between workers are made noisy with an error representing the recruiter's confidence levels. 
For the platform-based strategy, the error level is proportional to the history of workers (i.e., workers with more history in the platform have lower uncertainty levels). On the other hand, the uncertainty levels for the leader-based strategy are proportional and increase with the number of hops between the team leader and other workers in the social network. All the simulations were realized using the Monte Carlo method where 1000 realizations were made with different experimental settings. Additional details regarding the system model parameters and the optimization problem can be found in~\cite{hamrouni2020optimal}.}

\textcolor{black}{
In Fig.~\ref{4figures}, we illustrate the achieved metrics reflecting the team efficiency using the aforementioned recruitment strategies.} We notice that the leader-based strategy achieves the lowest team skill level and the highest relationship degree. This can be explained by the fact that the chosen leader recruits socially nearby workers and therefore, prioritizes the relationship levels over the team skill levels, which is completely opposite to the platform-based strategy. We also notice that the platform-based strategy has the highest recruiter uncertainly level and lowest team cost among the three evaluated strategies. Overall, we note a performance trade-off between the leader-based strategy and the platform-based strategy. The former recruits a team with higher relationship level and lower recruiter uncertainly but this comes at the expense of a higher cost and lower skill level compared to the latter. Since the hybrid strategy is a mixture of the remaining two strategies, it achieves intermediate performances.

\textcolor{black}{
In Fig.~\ref{perfor}, we compare three algorithms to form CMC teams for the platform-based strategy: i) ILP, ii) GA, and iii) PSO while evaluating the  five aforementioned metrics. The uncertainly levels of the skills and the social relations describe the differences between the real values and the values estimated by the recruiter of the workers' skill levels and social network relationship strength.} The result of this simulation shows that the optimal ILP achieves the highest performances for all the five metrics resulting in a high overall objective function. The GA algorithm recruits teams with higher skill levels, relationships levels, and lower cost and uncertainty. The approximation factor between the overall performances of the CMC GA and the optimal ILP does not bypass $1.15$. In fact, the genetic algorithm was able to recruit teams that have close metrics to the optimal ILP. The PSO achieves close but lower performances than the~GA. Devising low complexity algorithms may enable the real-time recruitment of teams in CMC.

\section{Perspective and Future Research Directions}
\label{prespective}
In this section, we discuss the challenges and perspective of CMC and highlight several directions for future research.

$\bullet$ \textbf{Spatial Tasks Integration and Privacy:} Most existing CMC approaches consider mobile crowdsourcing tasks but do not take into account spatial tasks (i.e., workers are asked to travel to tasks' locations). This Spatial Collaborative MC (SCMC) is a challenge because the team formation process can be dynamic (i.e., teams change when traveling from one task location to another) and complex because not only it considers the space dimension but also the time dimension (i.e., tasks are also tagged with beginning and ending time). Also, the server task assignment model includes the collection of the location information of all potential volunteers, which may reveal workers’ locations, and consequently raises privacy issues.

$\bullet$ \textbf{Server-Worker Interaction and Coordination:} Most existing task assignment models focus on the server assigning tasks, which assume that workers should perform the tasks once they receive the assignments. However, in practice, it is possible that workers may refuse to complete the tasks due to various reasons (e.g., laboriousness). Hence, it is more efficient and effective to combine the server assignment model with a worker selection model so that workers may select the projects they are good at and then compete with others to obtain the reward. Also, the CMC server needs to be instantly updated about the projects' status. It must keep track of the teams' progress and coordinate their work and projects' schedules.

$\bullet$ \textbf{Social Internet-of-Things for Large-scale CMC:}
Naturally, the requester prefers trustworthy and reliable workers to execute the different tasks of the project. Therefore, the social Internet-of-Things (SIoT) concept~\cite{MARCHE2020107248}, where IoT objects can establish social relations, can be exploited in CMC to leverage large-scale recruitment of IoT workers, ensure high-level of trustworthy and reliable operation, and cope with privacy and security concerns. The social relationships are established between various IoT objects (e.g., smart-phones, autonomous vehicles) and are built according to several criteria such as the communication links, locations, owners' policies, and interactions between objects. Moreover, SIoT can promote the use of community detection and clustering algorithms to help reduce the complexity and time of the recruitment and task assignment processes in CMC platforms.

$\bullet$ \textbf{Power Consumption and Traffic Load:}
\textcolor{black}{Hiring a team of IoT workers to collaborate together to complete projects while monitoring their progress requires a fair amount of energy and induces additional traffic load on the network. Mobile energy profiling, which is characterizing the energy consumption of a mobile device including installed applications, hardware, and other subsystem components, might be required in order to predict and eventually reduce the unnecessary energy utilization. 
The amount of energy consumption can vary from one CMC application to another depending on which sensors or resources are needed. Also monitoring the teams' progress can cause some energy load (e.g., tracking locations of the workers using their GPS sensor). }

\section{Conclusion}
\label{conclusion}
CMC is a state-of-the-art tangled paradigm especially for large-scale IoT. Unlike typical MC, CMC must consolidate the social aspects to successfully accomplish crowd tasks. This can be achieved by carefully designing the CMC projects, devising robust incentive mechanisms, adopting effective team recruitment strategies, and effectively monitoring the quality of responses. The paradigm has not yet reached maturity and a huge amount of work must be undertaken on several fronts to enable scalable, secure, and sustainable CMC systems.

\bibliographystyle{IEEEtran}
\bibliography{references}

\begin{IEEEbiography}
[{\vspace{0.1cm}\includegraphics[width=1.1in,height=1.34in]{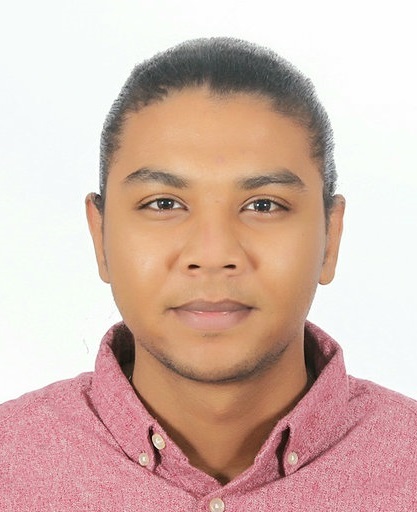}}]
{AYMEN HAMROUNI}
(Student Member, IEEE) received the Diplome d’Ingenieur degree (\textit{summa cum laude}) in telecommunication engineering from the Ecole Superieure des Communications de Tunis (SUP’COM), Tunis, Tunisia, in 2019. In 2019, he worked as a Research Assistant with the Stevens Institute of Technology, Hoboken, NJ, USA. He is currently a Research Scholar with the School of Systems and Enterprises, Stevens Institute of Technology. His research interests include the intersection of mobile crowdsourcing, applied machine learning, optimization, network analysis, mathematical modeling, graph theory, and the Internet-of-Things.
\end{IEEEbiography}
\vskip -5pt plus -1fil
\begin{IEEEbiography}
[{\vspace{0.1cm}\includegraphics[width=1.1in,height=1.34in]{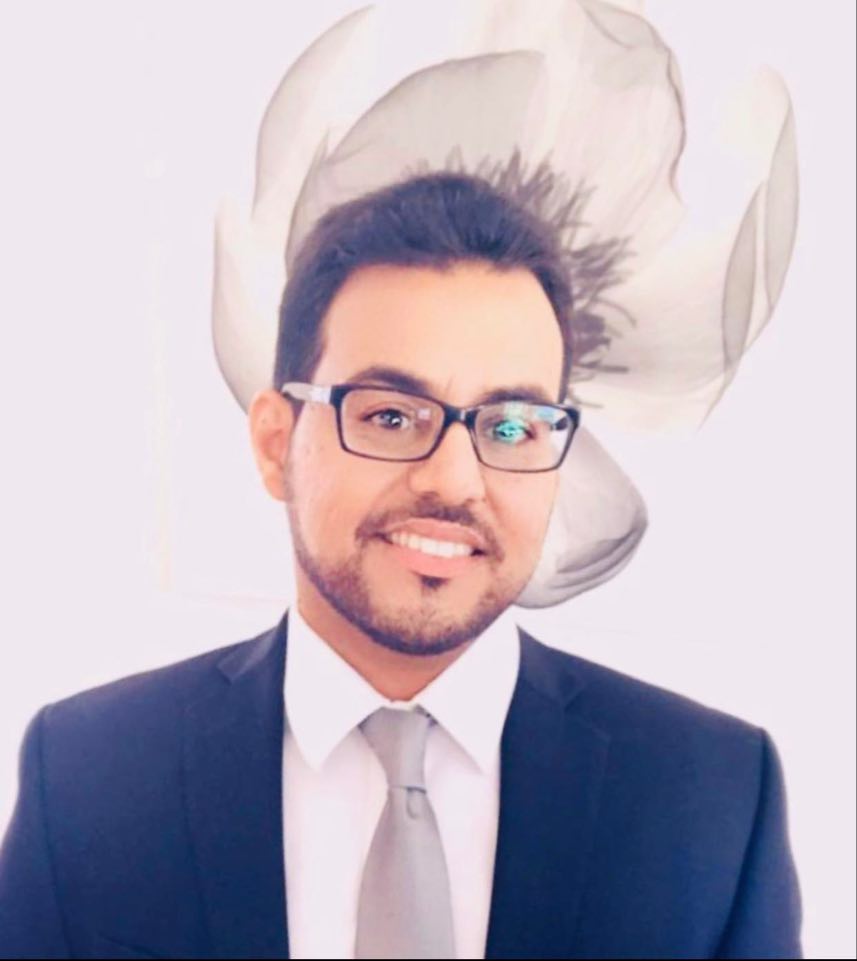}}]%
{TURKI ALELYANI} (Member, IEEE) received his Ph.D. in Software Engineering from Stevens Institute of Technology, New Jersey, United States. Prior to that, he received his master degree in Computer Science from Stevens as well. Dr. Alelyani's research studies Socio-Technical Systems Design in order to overcome some of the challenges in motivation, engagement, coordination and collaboration. His research is applied into different domains including engineering design, Software Engineeing, and social computing. He approaches this research by conducting empirical studies which involve using statistical analysis, machine learning, and experimental design.

\end{IEEEbiography}
\vskip -5pt plus -1fil
\begin{IEEEbiography}
[{\vspace{0.1cm}\includegraphics[width=1.1in,height=1.34in]{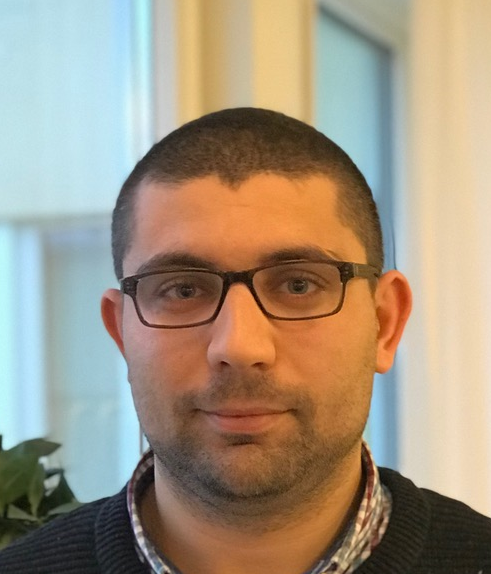}}]
{HAKIM GHAZZAI} (Senior Member, IEEE) is currently working as a research scientist at Stevens Institute of Technology, Hoboken, NJ, USA. He received his PhD degree in Electrical Engineering from KAUST in Saudi Arabia in 2015. He received his Diplome d’Ingenieur and Master degree in telecommunications from the Ecole Superieure des Communications de Tunis (SUP’COM), Tunis, Tunisia in 2010 and 2011. Before joining Stevens, he worked as a visiting researcher Karlstad University, Sweden and as a research scientist at Qatar Mobility Innovations Center (QMIC), Doha, Qatar from 2015 to 2018.  His general research interests include wireless networks, UAVs, Internet-of-things, and intelligent transportation systems.
\end{IEEEbiography}
\vskip -5pt plus -1fil
\begin{IEEEbiography}
[{\includegraphics[width=1.1in,height=1.34in]{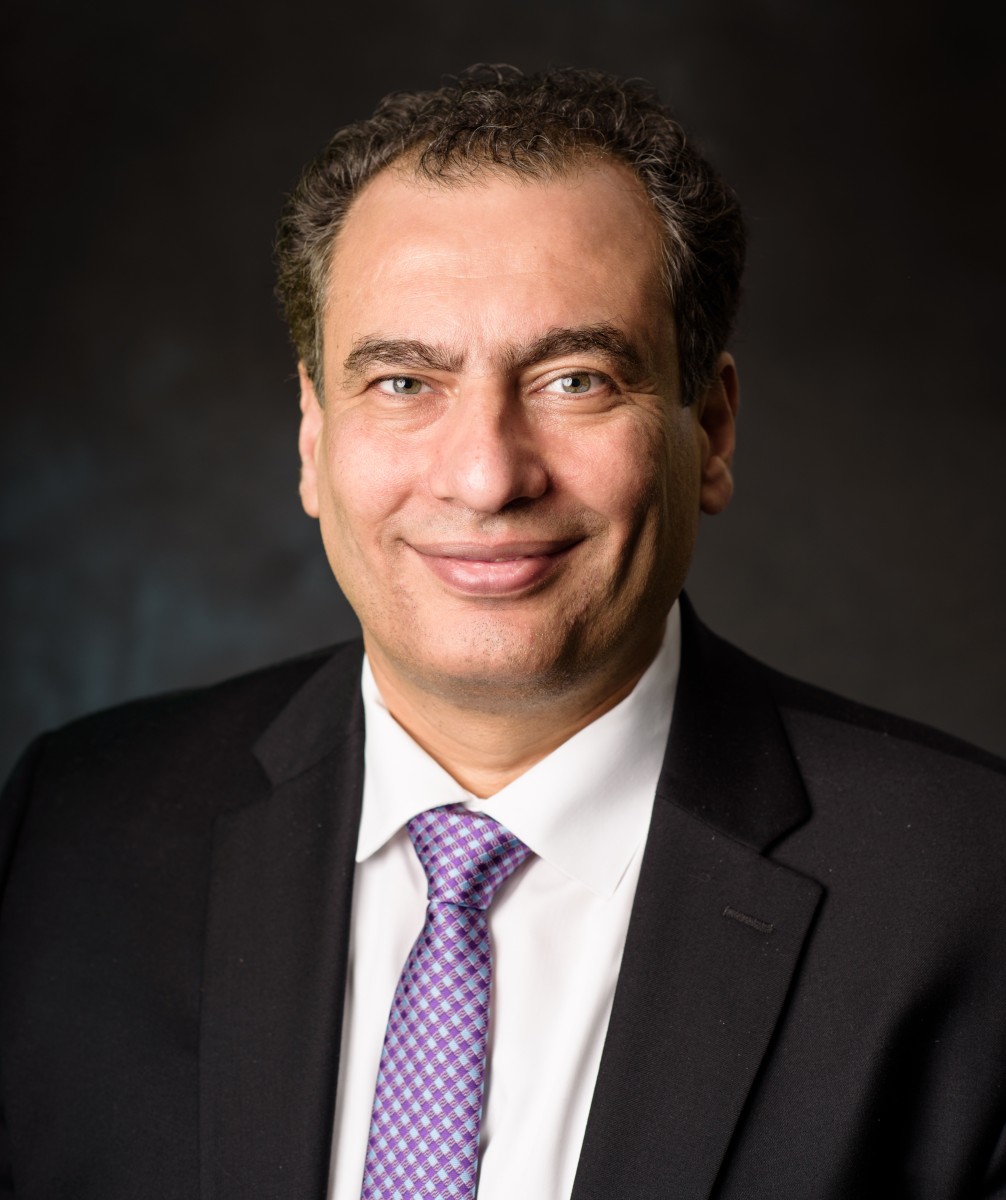}}]
{YEHIA MASSOUD} (Fellow Member, IEEE) received the Ph.D. degree from the Massachusetts Institute of Technology, Cambridge, USA. He is currently the Dean of the School of Systems and Enterprises, Stevens University of Science and Technology, USA. He is a Fellow of the IEEE and he has authored over 300 articles in peer-reviewed journals and conferences. He was selected as one of ten MIT Alumni Featured by MIT’s Electrical Engineering and Computer Science department in 2012. He was a recipient of the Rising Star of Texas Medal, the National Science Foundation CAREER Award, the DAC Fellowship, the Synopsys Special Recognition Engineering Award, and two best paper awards. Dr. Massoud has held several academic and industrial positions, including a member of the technical staff with Synopsys, Inc., CA, USA, a tenured faculty with the Departments of Electrical and Computer Engineering and Computer Science, Rice University, Houston, USA, and the Head of the Department of Electrical and Computer Engineering, Worcester Polytechnic Institute, USA. Massoud has served as the editor of Mixed\textendash Signal Letters\textendash The Americas and also as an associate editor of IEEE Transactions on Very Large Scale Integration Systems and IEEE Transactions on Circuits and Systems I.
\end{IEEEbiography}
\vfill
\end{document}